\documentclass{INTERSPEECH2023}
\usepackage{graphicx}
\usepackage{subfigure}
\usepackage{float}
\usepackage{cite}
\usepackage{multicol}

\interspeechcameraready

\title{What is Learnt by the LEArnable Front-end (LEAF)? Adapting Per-Channel Energy Normalisation (PCEN) to Noisy Conditions}
\name{Hanyu Meng, Vidhyasaharan Sethu, Eliathamby Ambikairajah}

\address{
  University of New South Wales, Australia}
\email{\{hanyu.meng, v.sethu, e.ambikairajah\}@unsw.edu.au}
\usepackage{multicol}

\begin{document}
\maketitle
\begin{abstract}
There is increasing interest in the use of the LEArnable Front-end (LEAF) in a variety of speech processing systems. However, there is a dearth of analyses of what is actually learnt and the relative importance of training the different components of the front-end. In this paper, we investigate this question on keyword spotting, speech-based emotion recognition and language identification tasks and find that the filters for spectral decomposition and the low pass filter used to estimate spectral energy variations exhibit no learning and the per-channel energy normalisation (PCEN) is the key component that is learnt. Following this, we explore the potential of adapting only the PCEN layer with a small amount of noisy data to enable it to learn appropriate dynamic range compression that better suits the noise conditions. This in turn enables a system trained on clean speech to work more accurately on noisy test data as demonstrated by the experimental results reported in this paper.

\end{abstract}

\section{Introduction}
The speech front-end is a crucial component in speech signal classification systems, and has been the focus of research for many decades. The Mel filterbank~\cite{mel} leads to representations inspired by our understanding of human perception, and is arguably the most widely used front-end across a range of different speech applications. With the advent of deep learning systems, there has been an interest in end-to-end systems that attempt to learn the optimal transformations to extract information from speech waveforms for any target application~\cite{tasnet,e2e_model_comparison,towards_e2e,end_to_end_ASV}. More recently, focus has shifted to learnable front-ends that constraints the architecture of the front-end but allows the model parameters to be learnt in conjunction with the back-end~\cite{zeghidour2021leaf,learnable_speaker_verification,learnable_emotion_recognition, fastaudio,sincnet,cgcnn,td_fbank}. They have been employed in a wide range of applications, including but not limited to speaker verification~\cite{learnable_speaker_verification}, spoofing detection~\cite{fastaudio}, and emotion recognition~\cite{learnable_emotion_recognition}.

A number of general-purposed learnable front-ends such as Time-Domain Filterbank (TD-Fbank)~\cite{td_fbank}, SincNet~\cite{sincnet}, CGCNN~\cite{cgcnn} and LEArnable Front-end (LEAF)~\cite{zeghidour2021leaf} have been developed recently. Among these, LEAF stands out as having fewer parameters and higher reported accuracy in tasks such as audio events classification. Its universal representation generated from raw speech signals has also made it applicable in other audio-related tasks, such as medical acoustic signal feature learning~\cite{leaf_in_medical}, analog acoustic recognition~\cite{peaf}, bird activity detection~\cite{leaf_in_bird_activity}, speaker verification~\cite{leaf_speaker_verification}, and limited-vocabulary speech recognition tasks~\cite{leaf_different_dataset}. Despite these successes, little is known about what is exactly learnt by LEAF from speech signals.

Analyses of the LEAF model have suggested that one of its components, Per-Channel Energy Normalization (PCEN) plays an important role in effectively compensating for the impact of environmental noise on speech intelligibility~\cite{pcen_why_how,pcen_sed,pcen_noise_1,pcen_noise_2}. PCEN has also been widely applied in acoustic scene classification~\cite{acoustic_scene_pcen} and long-distance bioacoustic event detection~\cite{long_distance_detection_pcen}. However, beyond this, there have been limited analyses and insights in the operation of LEAF. To address this shortcoming, we investigate which components of LEAF learn during model training and to what extent.

In this paper, we demonstrate that only the PCEN layer of the broader LEAF model learns during training and there is no observable change in the characteristics of any of the other components away from their initial values. Following this, we leverage our findings to develop a noise adaptation strategy whereby only the PCEN layer of LEAF is adapted using a small amount of noisy data to enable the LEAF model to be used under noisy conditions.

\section{What is learnt by LEAF?}
\subsection{LEAF}
\begin{figure}[ht!]
  \centering
  \includegraphics[width=\linewidth]{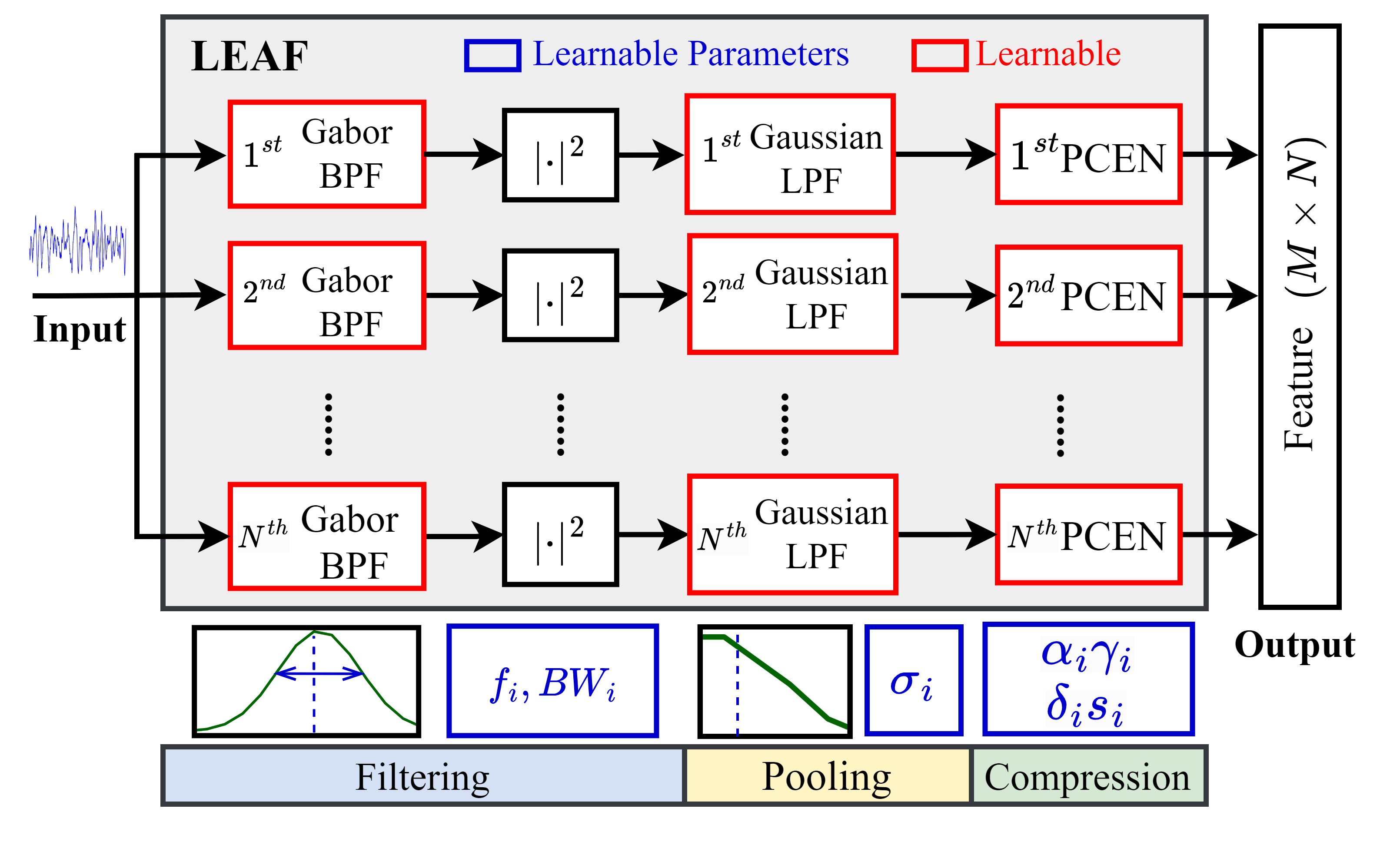}
  \caption{An overview of LEAF (the input speech frame contains $M$ samples).}
  \label{fig1:LEAF}
\end{figure}

LEAF is a general-purpose audio front-end designed for audio event classification~\cite{zeghidour2021leaf}. It mainly comprises three learnable and one non-learnable elements: Spectral Decomposition, Energy Estimation (non-learnable), Smoothing, and Dynamic Range Compression. As depicted in Figure~\ref{fig1:LEAF}, a frame of speech with $M$ samples passes through a parallel filterbank of $N$ Gabor filters~\cite{gabor}, which is initialised to be equally spaced along the mel-frequency scale. During training, both the centre frequencies $(f_i)$ and bandwidths $(BW_i)$ of all filters can be learnt. The filterbank is followed by energy estimation implemented by a sample-wise squaring operation, which in turn is smoothed using a low pass filter. The low pass filtering is implemented as a pooling operation comprising of a Gaussian Low-pass Filters (LPF), an approach that has been shown to be effective with 2D features~\cite{gaussian_pooling}. The learnable parameter is the standard deviation $\sigma_i$ (equivalently its bandwidth) of the Gaussian LPF in each frequency channel.

Finally, the per channel energy compression (PCEN) acts as an input energy level-dependent gain controller that computes appropriate gains to enhance or attenuate signals in each frequency channel~\cite{original_pcen}. The PCEN layer for the $i^{th}$ frequency channel takes as input the smoothed energy estimates produced by the low pass filter layer, $E[n,i]$, and is formulated as:

\begin{equation}
    PCEN[n,i] = \left(\frac{E[n,i]}{\left(M[n,i]+\epsilon\right)^{\alpha_i}}+\delta_i\right)^{\gamma_i} - \delta_i^{\gamma_i},
\label{pcen1}
\end{equation}
where $M[n,i]$ is the smoothed version of the input representation achieved by a first-order infinite impulse response (IIR) filter as expressed in Equation~\ref{pcen2} with a learnable smoothing factor, $s_i$. 
\begin{equation}
    M[n,i] = s_iE[n,i] + (1-s_i)M[n-1,i].
    \label{pcen2}
\end{equation}
As illustrated in Figure~\ref{fig1:LEAF}, the PCEN layer has four learnable parameters $(s_i, \alpha_i, \delta_i, \gamma_i)$ per frequency channel. 
\subsection{Analysis of LEAF Parameters: Experimental Setup}
To investigate what is learnt by LEAF, we train LEAF on three of the tasks where its performance has been perviously reported: emotion recognition, keyword spotting, and language identification~\cite{zeghidour2021leaf}. The tasks were chosen to represent a range of speech processing applications as well as prediction accuracies ranging from quite high to somewhat low (refer Table ~\ref{accuracy_table}). Following this, we compare the spectral and compression characteristics of the three learnable components of LEAF before and after training.

For these analyses, we replicate the dataset settings reported in~\cite{efficientleaf}. The keyword spotting task is trained on SpeechCommandsV2 dataset~\cite{speechcommands}, emotion recognition experiments were carried out on CREMA-D dataset~\cite{crema_d}, and we use Voxforge~\cite{voxforge} for language identification. All three datasets are sampled at 16kHz. For all three tasks, we utilise EfficentNetB0~\cite{efficientnet} as the back-end, which is a lightweight Convolution Neural Network (CNN) based classification network.

Prior to training, the LEAF model was initialised as described in~\cite{zeghidour2021leaf}. Specifically, we initialised the 40 Gabor filters using the mel-scale and set the input window size to 401 samples with a hop size of 160 samples, corresponding to 25ms for audio sampled at 16kHz. The Gaussian LPFs were initialised with a standard deviation of 0.4 for all channels. The initial values of the PCEN layer parameters in each frequency channel were set to $\alpha = 0.96$, $\gamma = 2$, $\delta = 2$, and $s = 0.04$.

For training the models, we utilised the ADAM optimiser with a fixed learning rate of $10^{-4}$, and employed mini-batches of size 256. We set the input sequence length for SpeechCommandsV2 and Voxforge to 1 second, and for CREMA-D as 3 seconds, based on the audio files durations in each dataset. To ensure consistent loudness range across different recordings, we rescaled the raw speech signals to a range between 15 dB Sound Pressure Level (SPL) and 30 dB SPL. In the test phase, we adopted the approach in ~\cite{efficientleaf}, and compute predictions for non-overlapping one-second segments and averaging the logits across the entire recording.

Table~\ref{accuracy_table} presents the test accuracy for all tasks with four different LEAF model settings. The ``Untrained'' settings indicate that all parts of LEAF were frozen and the initial values were not updated during training. The ``Filter Trained'' settings indicate that Gabor filters and Gaussian low pass filters were set to be trainable, and their parameters would be updated during model training (but the PCEN layer parameters would not be updated). Conversly, the ``PCEN Trained'' settings refer to the condition where only the PCEN layer parameters were trained and the Gabor filter and Gaussian low pass filter parameters were kept unchanged from their initial values. Finally, ``Fully Trained'' settings refer to the standard setting where all parameters are trainable.

From the table, the first interesting observation that stands out is that there is little difference between the various trained and untrained versions of LEAF across all three tasks. It is worth noting that for the emotion recognition task, where the untrained LEAF shows the highest accuracy, we used the data partitioning reported in~\cite{efficientleaf}, whereby the data was partitioned by shuffling speaker groups and segmenting the data to ensure speaker independence in each partition.  For the other two speech tasks, we used the data partitions as per the original dataset release. These results prompt the question \emph{``What is learnt by the LEAF model?''}. Specifically, through our analyses, we aim to answer the question: Which element of the model has undergone the most significant changes as a result of the training?

\begin{table}[ht!]
\caption{Classification accuracy of LEAF models (mean $\pm$ std.dev, over three runs).}
\label{accuracy_table}
\begin{tabular}{lllll}
\toprule
& \begin{tabular}[c]{@{}l@{}}Keyword\\ Spotting\end{tabular} & 
\begin{tabular}[c]{@{}l@{}}Emotion\\ Recognition\end{tabular} & \begin{tabular}[c]{@{}l@{}}Language\\ Identification\end{tabular} \\ \midrule
\begin{tabular}[c]{@{}c@{}}Untrained\end{tabular}                  & 94.78 ± 0.1                                                & \textbf{44.01±8.6}                                                     & 91.73±0.4                                             \\ 
\begin{tabular}[c]{@{}c@{}}PCEN Trained\end{tabular} & 95.07 ± 0.2                                                & 39.66±1.5                                                     & 89.95±2.1                                             \\ 
\begin{tabular}[c]{@{}c@{}}Filters Trained\end{tabular}                   & 94.62 ± 0.1                                                & 40.11±1.3                                                     & \textbf{95.13±1.4}                                            \\ 
\begin{tabular}[c]{@{}c@{}}Fully Trained\end{tabular}          & \textbf{95.18 ± 0.3}                                                & 41.10±1.4                                                      & 91.03±0.6                                             \\ \bottomrule
\end{tabular}
\end{table}

\begin{figure*}[!ht]
\centering
  \subfigure[Gabor filters' centre frequency changes]{
\includegraphics[width=.32\textwidth]{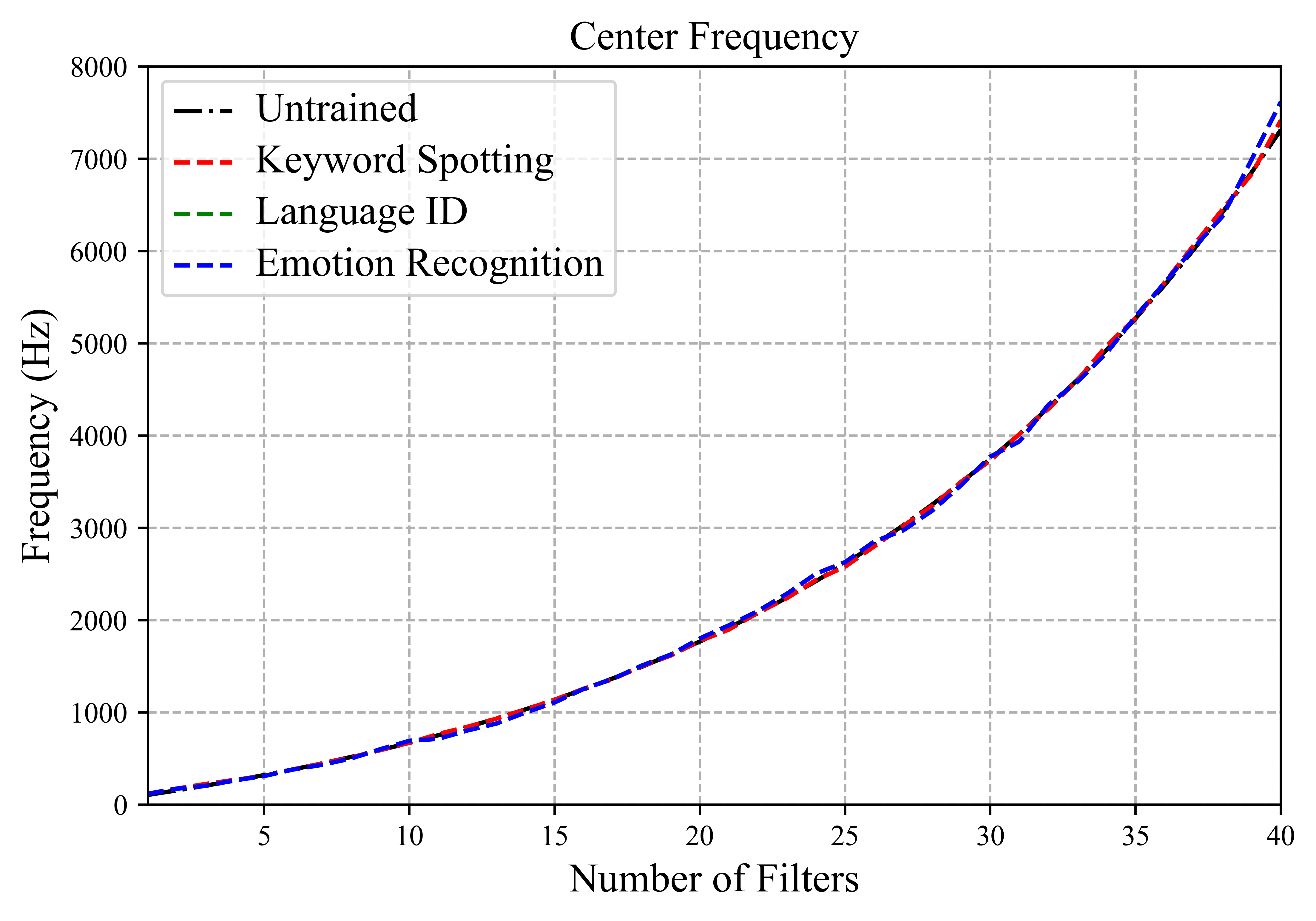}\label{fig:center_frequency}}
    \hfill
  \subfigure[Gabor filters' bandwidth changes]{
    \includegraphics[width=.32\textwidth]{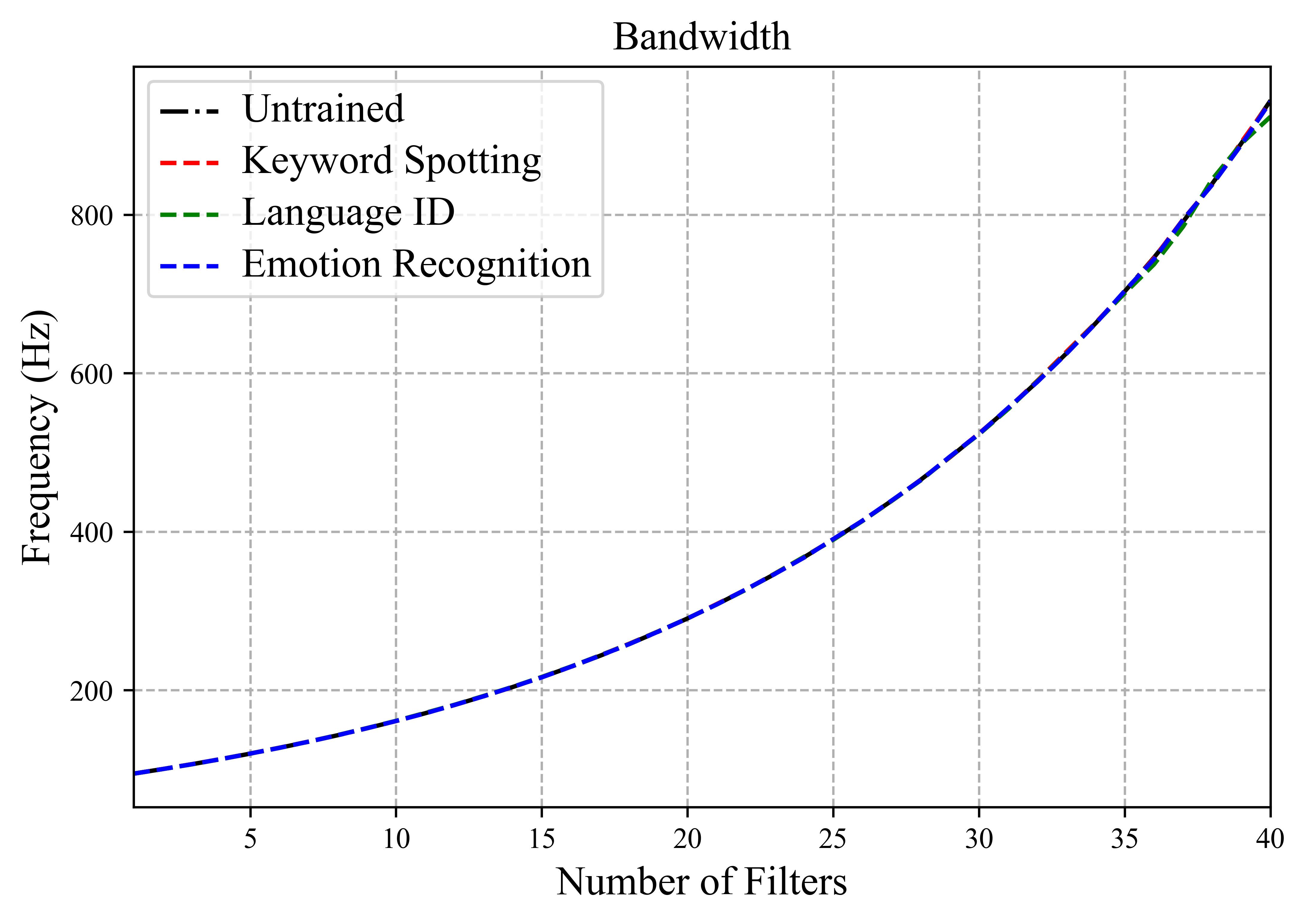}\label{fig:bandwidth}}
    \hfill
  \subfigure[Gaussian frequency response changes]{%
    \includegraphics[width=.32\textwidth]{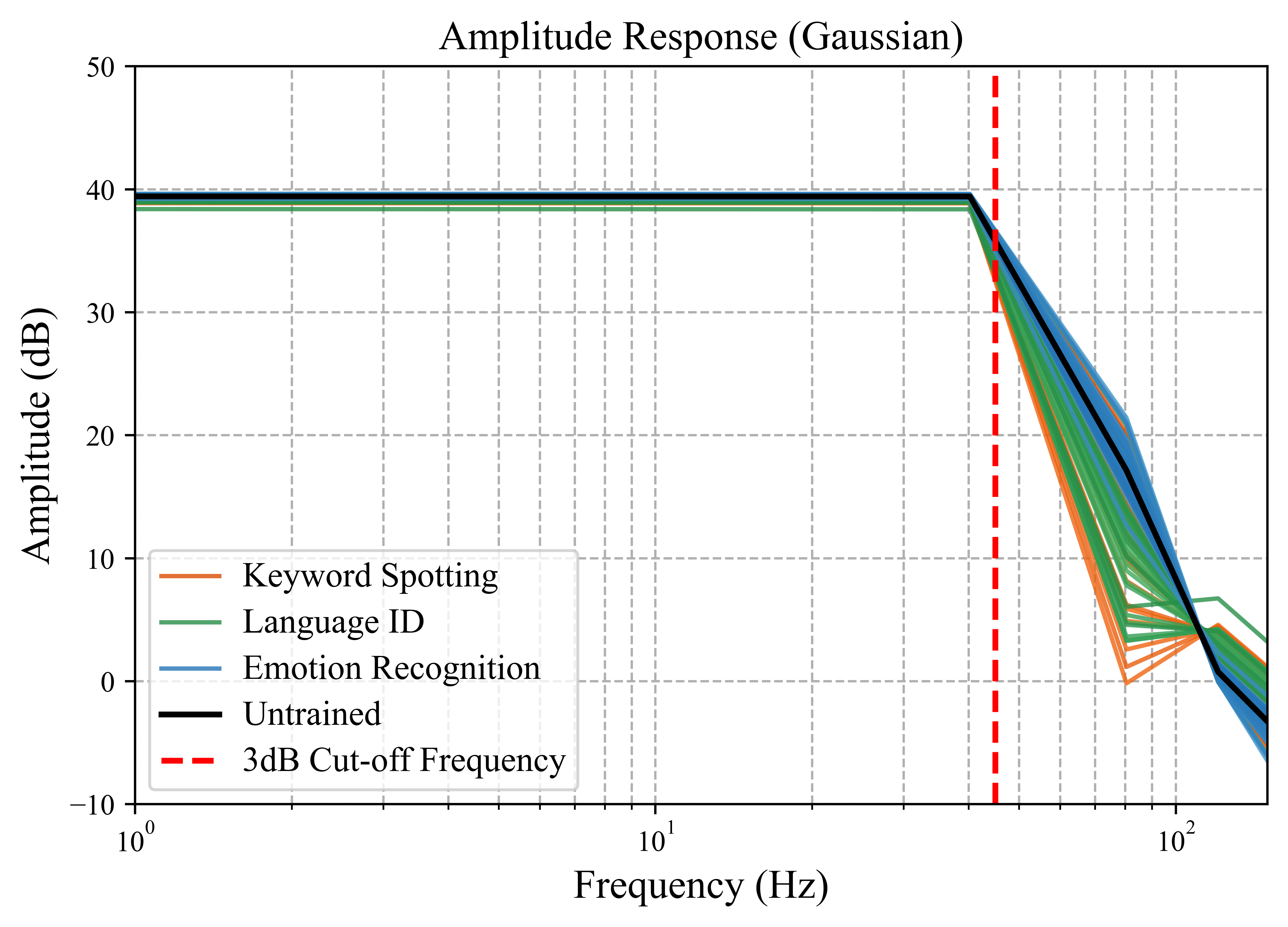}\label{fig:gaussian_3db}}
    \hfill
  \subfigure[PCEN gain changes (Keyword Spotting)]{%
    \includegraphics[width=.32\textwidth]{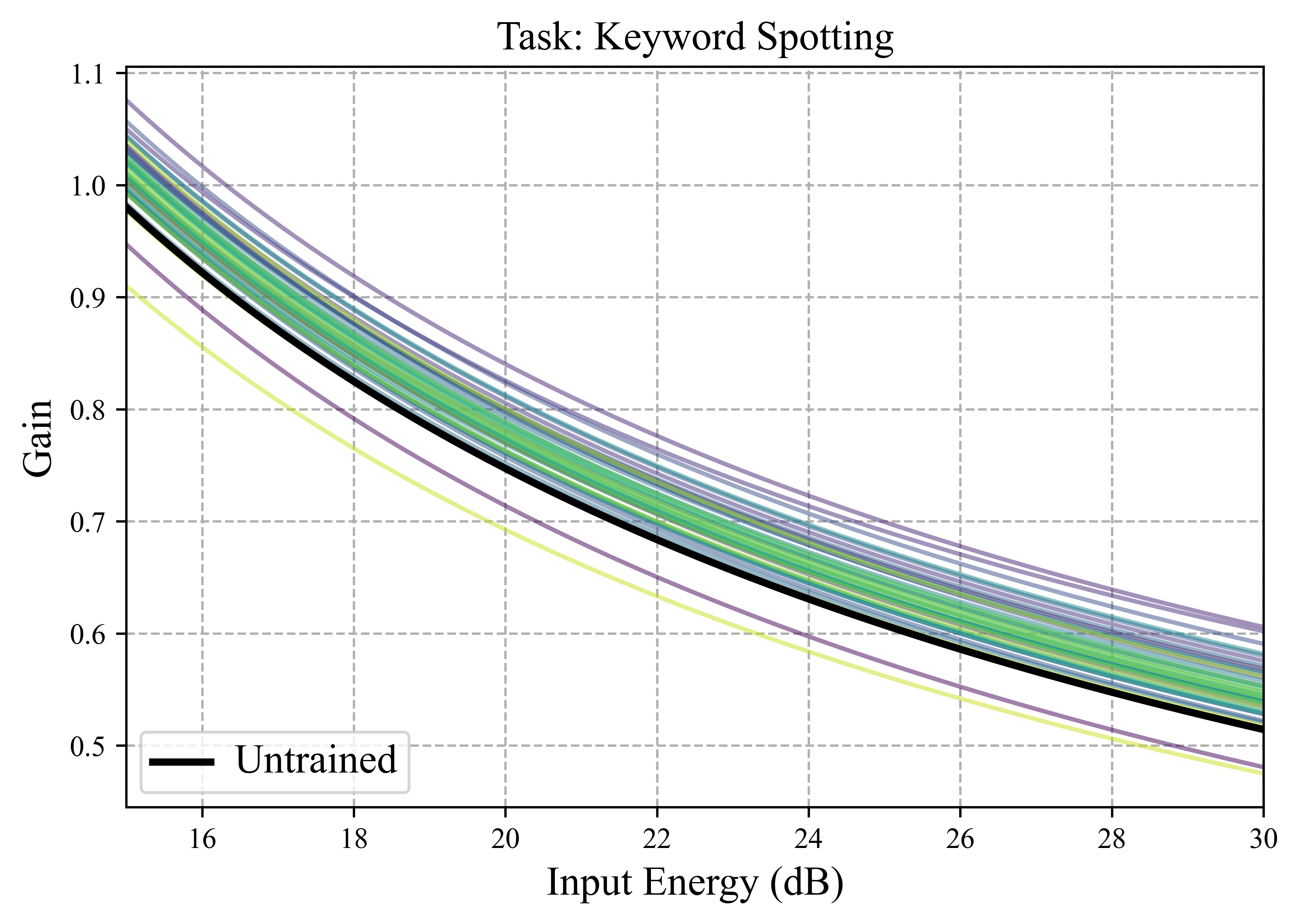}\label{fig:keyword_spotting}}
    \hfill
  \subfigure[PCEN gain changes (Emotion Recognition)]{%
    \includegraphics[width=.32\textwidth]{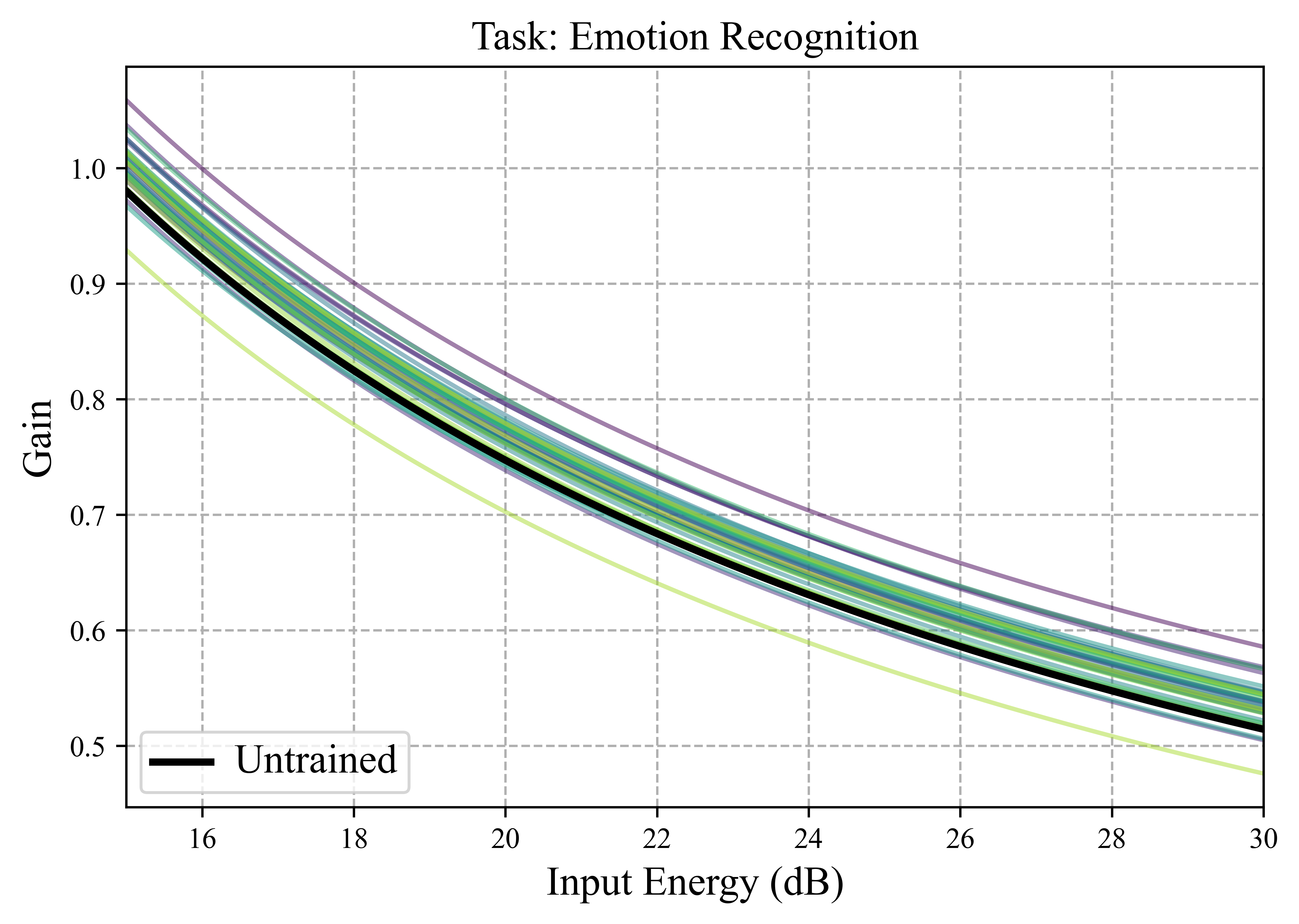}\label{fig:emotion_recognition}}
    \hfill
  \subfigure[PCEN gain changes (Language ID)]{%
    \includegraphics[width=.32\textwidth]{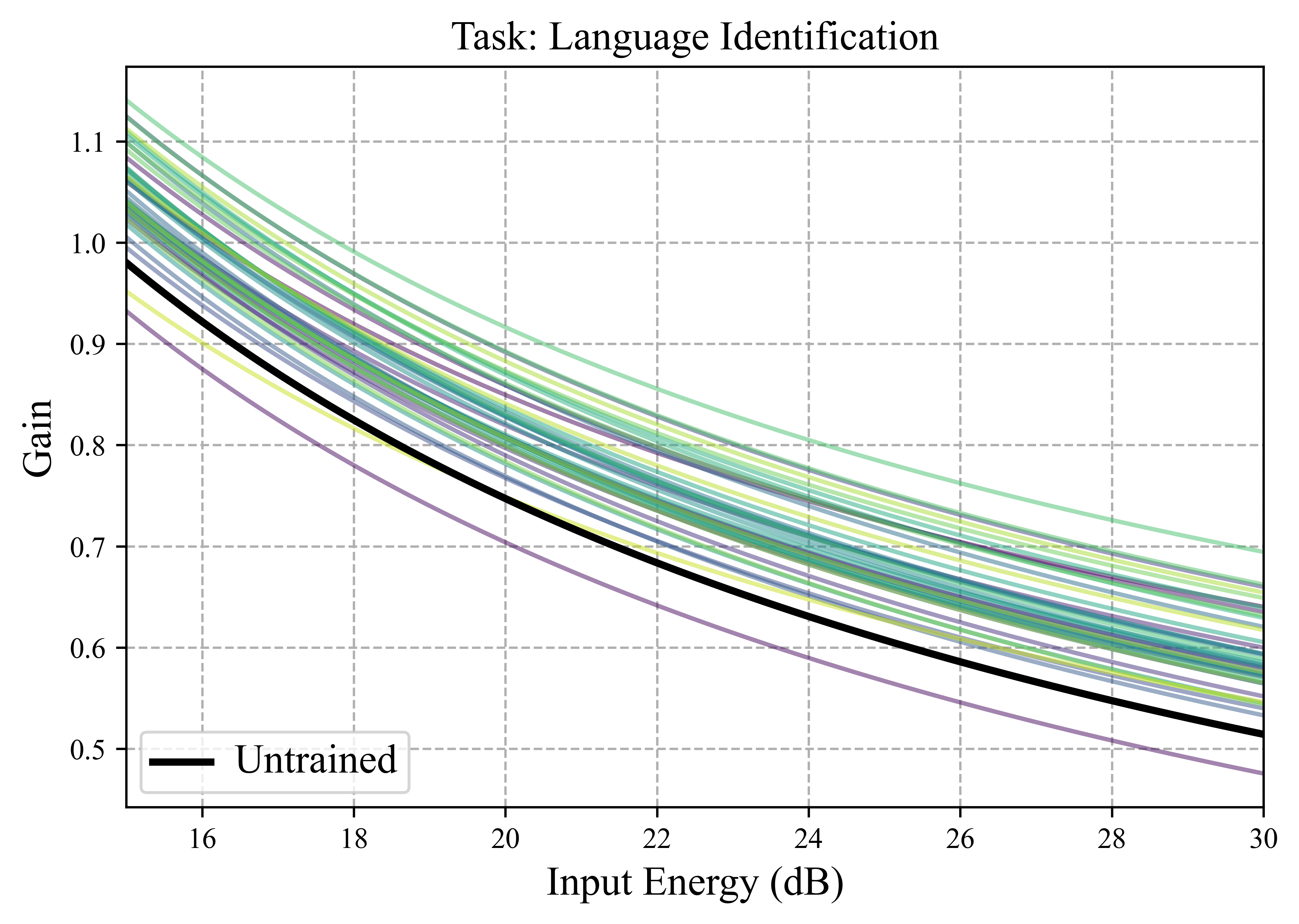}\label{fig:language_id}}
  \caption{Visualising learnt information of LEAF across three tasks.} 
  \vspace{-2mm}
\end{figure*}

\subsection{Analysis of LEAF Parameters: Results}
Extracting the weights from the fully trained and untrained LEAF models, we compute the centre frequencies and bandwidths of the Gabor filters as well as magnitude responses of the Gaussian low pass filters and compare them to each other (trained vs untrained across all three tasks). We also reconstruct the compression function of the PCEN layers and compare them to each other. These comparisons are shown in Figure~\ref{fig:center_frequency}-\ref{fig:language_id}.

\textbf{a) Gabor Filterbank:} Figures~\ref{fig:center_frequency} and~\ref{fig:bandwidth} represent the changes in centre frequencies and bandwidths of each filter of the 40 Gabor filters across three tasks as well as the initial values.  No appreciable deviations from the intial values can be observed for any of the trained models in any of the three tasks. These results strongly indicate that the initial Gabor filters may be optimal and learning does not help. 

\textbf{b) Gaussian Lowpass Filterbank:} In Figure~\ref{fig:gaussian_3db}, we plot the frequency response of all $40$ Gaussian lowpass filters for all three tasks. Once again, there appears to be no appreciable deviation from the initial values in the pass band of the low pass filters. This again suggests there may be no benefit to learning these filters.

\textbf{c) PCEN Range Compression:} In Figures~\ref{fig:emotion_recognition} to~\ref{fig:language_id}, we present input level dependent gain imparted by the PCEN layer across all frequency channels in each of the three tasks. To visualise this, we reconstruct the trained and untrained PCEN functions from Equations~\ref{pcen1} and~\ref{pcen2} and plot the PCEN gains as a function of input energy $E[n,i]$. The curves in Figures~\ref{fig:emotion_recognition} to~\ref{fig:language_id} transition from light green (for the $1^{st}$ channel) to dark purple (for the $40^{th}$ channel), represent the learnt gain characteristics for all 40 channels across the three tasks. It can be seen that the gain curves for the learnt models differ from that of the untrained model across all three tasks.

Taken together these results suggest that of three potentially learnable components of the LEAF model, the Gabor filterbank for spectral decomposition, the Gaussian low pass filter for energy smoothing and the PCEN which offers dynamic range compression, only the PCEN layers appear to be actually learning anything. Consequently, constraining the learning to only this layer of the LEAF model would significantly cut down the number of trainable parameters in the model, leading to more efficient learning. This in turn suggests a noise adaptation scheme involving a small set of noisy speech samples might enable a LEAF model trained on clean speech to be employed in noisy conditions. This hypothesis is explored in the rest of the paper.
\vspace{-2mm}
\section{Adapting PCEN to Noisy Environments}
We explore the hypothesis that adapting or tuning the PCEN layer in LEAF can enhance the accuracy of speech classification in noisy environments using a limited amount of noisy adaptation data. To test this hypothesis, we compare the performance of a system with a PCEN-adapted LEAF model to that of one trained on clean speech. For reference we also include a model trained entirely on a large amount of noisy data in the comparison\footnote{https://github.com/Hanyu-Meng/Adapting-LEAF}.

\subsection{Experiment Setups}
\subsubsection{Dataset and Partition}
Of the three speech procesing tasks, we choose the emotion recognition task to test the proposed PCEN adaptation scheme since it had the lowest accuracy. The CREMA-D dataset for speech emotion recognition consists of 91 speakers and 6 emotions~\cite{crema_d}, with each speaker having an almost equal number of utterances. When partitioning CREMA-D for this set of experiments, we use different partitions from those used in section 2.2. Specifically, we split the data based on sentences to ensure that each partition contained utterances from each speaker in order to minimise the impact of speaker variability on the results. As illustrated in Figure~\ref{fig:data_partitaion}, we used 9 sentences for training, 1 sentence for validation, and the remaining 2 sentences for testing. Consequently, the training set contained 5811 recordings, validation set contained 545, and the test set contained 1086. For adaptation, we selected one sentence from the training set (comprising of 546 recordings) as the adaptation data to which we add different types of noise (white noise and babble noise) at different Signal-to-Noise Ratios (SNRs).

\subsubsection{Experimental Setup}
\begin{figure}[ht!]
  \centering
  \includegraphics[width=0.7\linewidth]{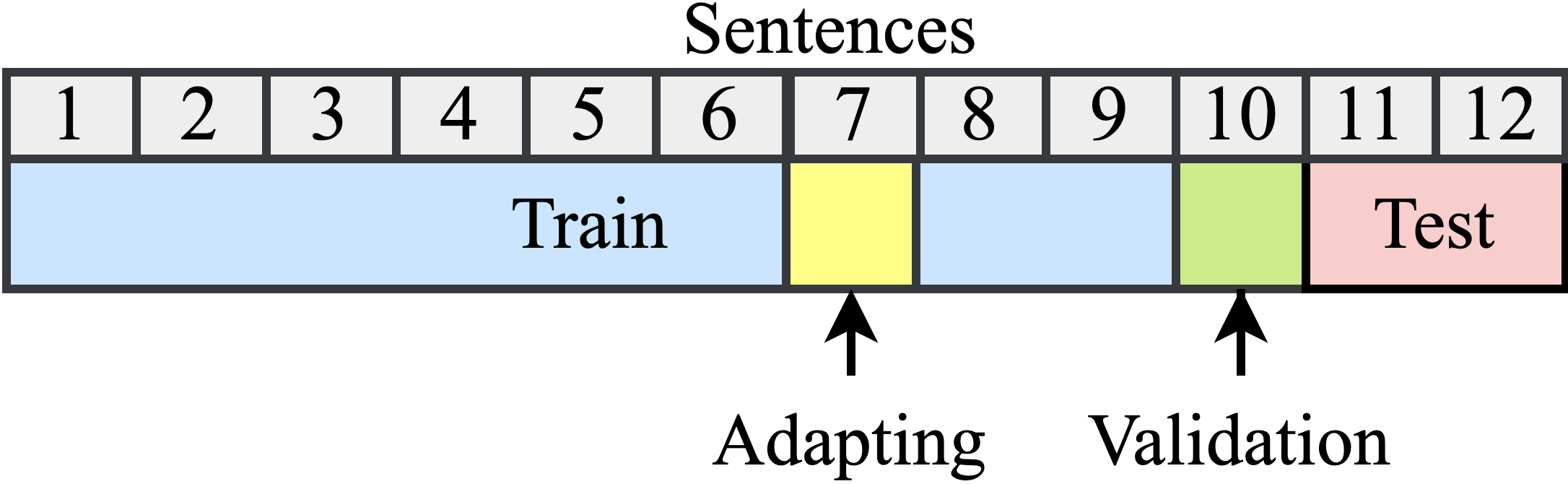}
  \caption{Data partitions in the CREMA-D dataset for PCEN adaptation experiments.}

  \label{fig:data_partitaion}
\end{figure}

\begin{figure}[ht!]
  \centering
  \includegraphics[width=\linewidth]{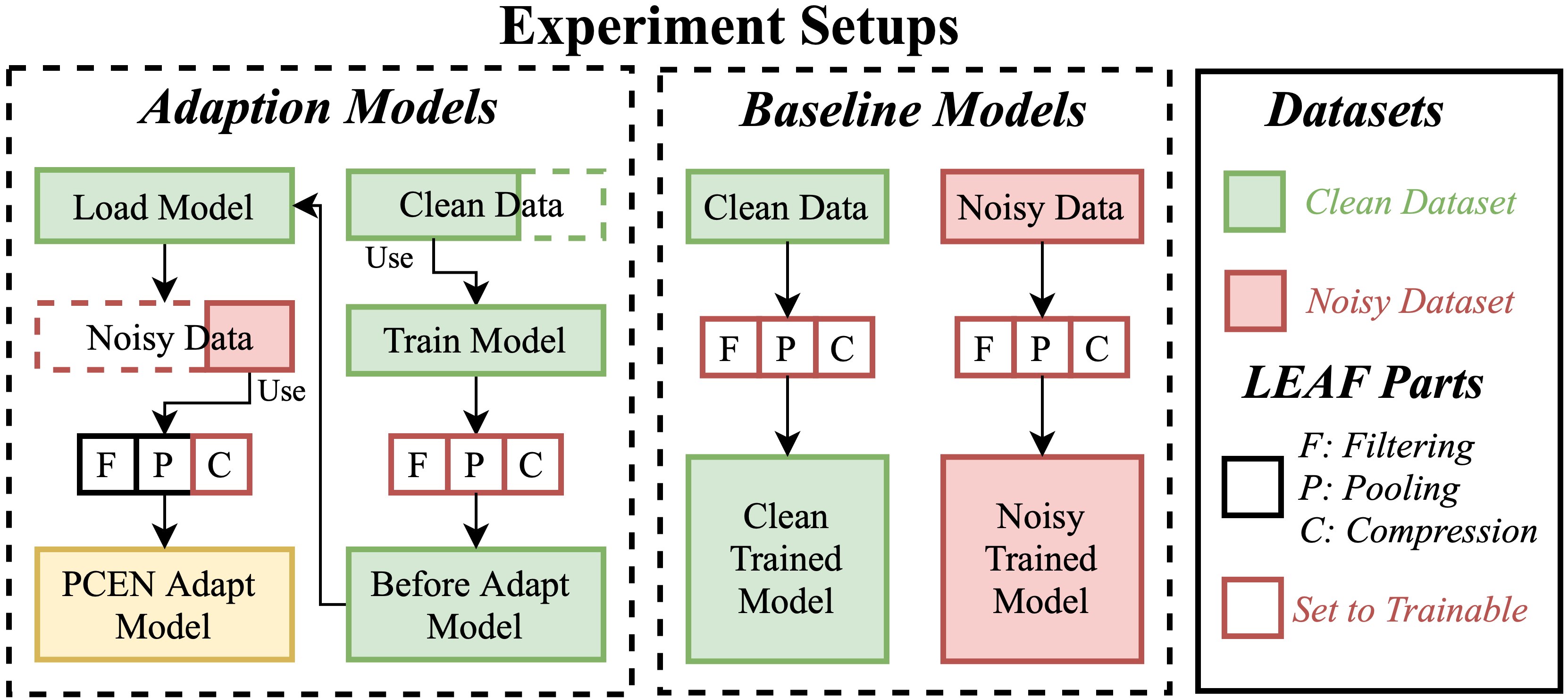}
  \caption{Overview of the PCEN adaptation experimental setup.}
  \label{fig:model_settings}
\end{figure}

For this experiment, we compared four models that had the same structure as the system used in the experiments reported in section 2.2, but with varying training setups as illustrated in Figure~\ref{fig:model_settings}. These models are:
\begin{itemize}
    \item \textbf{Clean Trained}: Trained on the entire noise-free training set and it serves as a baseline.
    \item \textbf{Noisy Trained}: Trained on a noisy version of the entire training data. 
    \item \textbf{Before Adaptation (BA)}: Trained on the noise-free training set without including adaptation data. This provides a reference level for performance prior to adaptation (see Figure~\ref{fig:data_partitaion}).
    \item \textbf{PCEN Adaptation (PA)}: This is the BA model with the PCEN layer adapted using the noisy adaptation data.
\end{itemize}

\subsection{Results and Analysis}
To verify the hypothesis proposed at the beginning of Section 3, we tested the adaptation of PCEN using both stationary and non-stationary noise.

\subsubsection{Gaussian Noise Adaptation}

The left pane in Figure~\ref{fig:noise_result} shows the classification accuracy for all four models tested by adding different levels of Gaussian noise to the clean test data in order to obtain SNR in the range of 0 dB to 20 dB. The results suggest that training the model with Gaussian noise helps the model learn the pattern of noise and improves its robustness. Further, the models trained on only clean data perform poorly when exposed to Gaussian noise. However, after adapting the PCEN layer with a small amount of noisy data, the impact of noise on accuracy can be somewhat mitigated as can be seen by comparing the performance Before Adaptation (BA model) to that after adaptation (PA model).

\subsubsection{Babble Noise Adaptation}

\begin{figure}[ht!]
  \centering
  \includegraphics[width=0.95\linewidth]{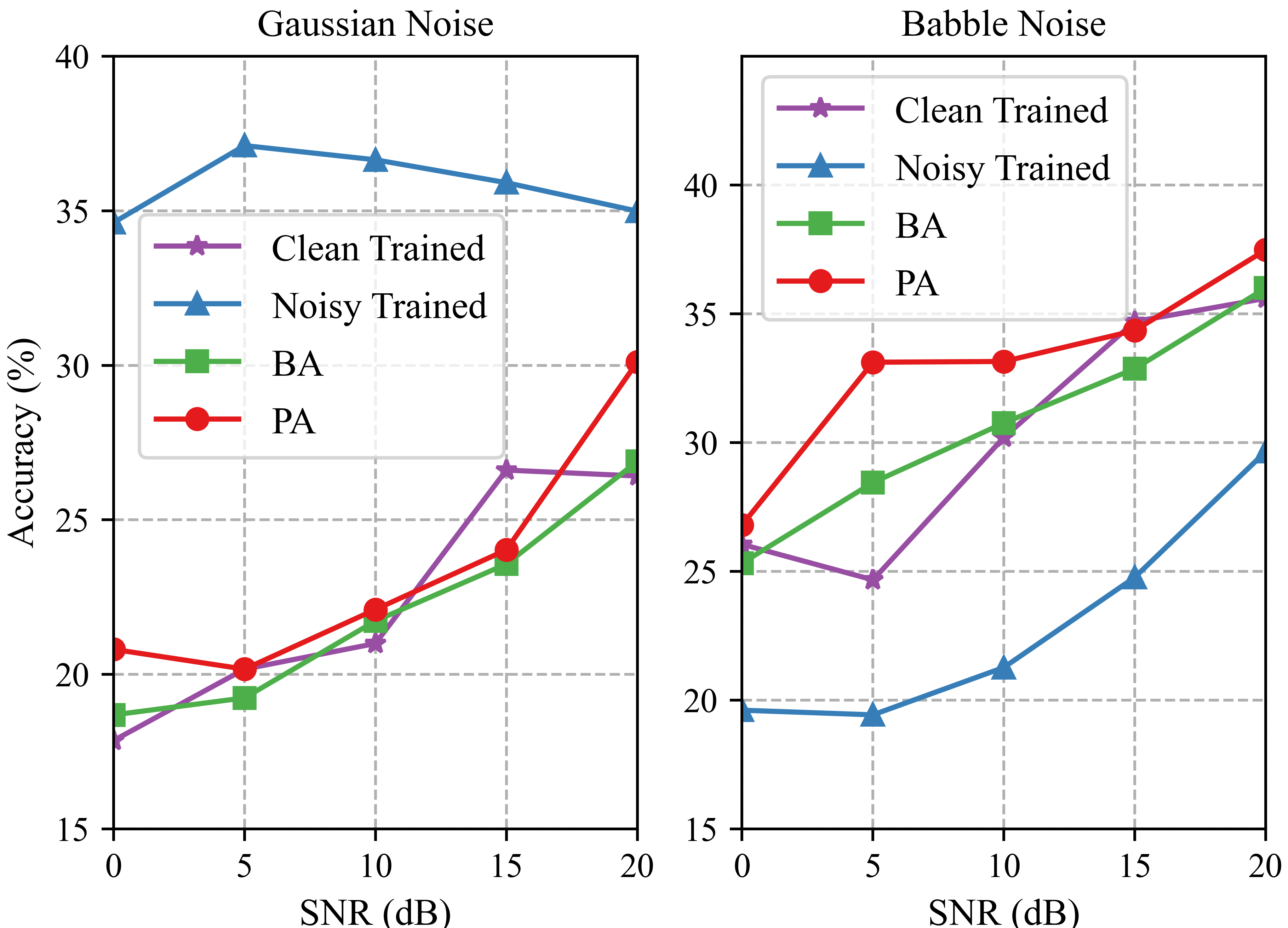}
  \caption{Accuracy of PCEN adaptation experiments on the CREMA-D dataset for different levels of Gaussian noise (left) and babble noise (right).}
  \label{fig:noise_result}
\end{figure}

We also repeated the above experiment using babble noise instead of white noise. To simulate babble noise, we followed the data augmentation approach used in~\cite{x_vector} and used the MUSAN dataset, which contains 60 hours of speech from 12 languages~\cite{musan}. Specifically, we randomly selected three speech recordings from MUSAN, mixed them together, and added them to the clean signal to simulate babble noise with SNR ranging from 0 dB to 20 dB. The right pane in Figure~\ref{fig:noise_result} shows how the accuracy of the four models changes under different levels of babble noise.

The results suggest that training with noisy data is not as effective under babble noise conditions. This might be due to the greater similarity between noise and speech in this case. However, we can observe from the graph that adapting PCEN with babble noise may be quite effective in allowing the model to be used under noisy conditions.

\section{Conclusion}
In this study, we sought to answer the question \emph{'What is learnt by LEAF?'}, by comparing the extent of change in the characteristics of the different learnable components of the LEAF model from their initial values prior to training. This analysis was repeated on multiple speech processing tasks, and consistently our analyses revealed that only the PCEN layer changes in response to training. The filterbank and low pass filters employed for spectral decomposition and spectral energy smoothing remained unchanged for all three tasks. This suggests that the actual learning in the LEArnable Front-end occurs within a much lower dimensional subspace of the parameter space of the model. Following this, we developed a model adaptation scheme constrained to this subspace (PCEN layer only) using a small amount of noisy training data to adapt a LEAF trained on clean speech to operate more effectively in noisy conditions.

\section{Acknowledgements}
This work was funded by ARC Discovery Grant DP210101228. The authors would also like to thank UNSW, Sydney, Australia for providing PhD scholarship support.
\bibliographystyle{IEEEtran}
\bibliography{mybib}

\end{document}